\documentclass[10pt,twocolumn,letterpaper]{article}

\usepackage[
]{utils/cvpr}      

\usepackage{float}
\usepackage{graphicx}
\usepackage{amsmath}
\usepackage{amssymb}
\usepackage{booktabs}
\usepackage{textcomp}
\usepackage{stfloats}
\usepackage{url}
\usepackage{svg}
\usepackage{verbatim}
\usepackage{multirow}
\usepackage{multicol}
\usepackage{array}
\usepackage{booktabs} 
\usepackage{pifont} 
\usepackage{xcolor}
\usepackage{amssymb}  
\usepackage[multiple]{footmisc}
\usepackage{bigfoot}

\usepackage[pagebackref,breaklinks,colorlinks]{hyperref}

\usepackage[capitalize]{cleveref}
\crefname{section}{Sec.}{Secs.}
\Crefname{section}{Section}{Sections}
\Crefname{table}{Table}{Tables}
\crefname{table}{Tab.}{Tabs.}
\def\cvprPaperID{7733} 
\def\confName{CVPR}
\def\confYear{2025}

\DeclareNewFootnote{AAffil}[arabic]
\DeclareNewFootnote{ANote}[fnsymbol]
\usepackage{etoolbox}
\makeatletter
\patchcmd\maketitle{\def\@makefnmark{\rlap{\@textsuperscript{\normalfont\@thefnmark}}}}{}{}{}
\makeatother

\makeatletter
\def\thanksAAffil#1{
  \footnotemarkAAffil\protected@xdef\@thanks{\@thanks%
        \protect\footnotetextAAffil[\the \c@footnoteAAffil]{#1}}%
}
\def\thanksANote#1{%
  \footnotemarkANote%
  \protected@xdef\@thanks{\@thanks%
        \protect\footnotetextANote[\the \c@footnoteANote]{#1}}%
}
\makeatother

\begin{document}

\title{Volume Tells: Dual Cycle-Consistent Diffusion for 3D Fluorescence Microscopy De-noising and Super-Resolution}


\author{
Zelin Li\thanksAAffil{Department of Electrical Engineering, City University of Hong Kong, Hong Kong (SAR China)}$^{,}$\thanksAAffil{Centre for Intelligent Multidimensional Data Analysis, Hong Kong Science Park, Hong Kong (SAR China)}, %
\and
Chenwei Wang\footnotemarkAAffil[2]$^{,}$%
  \thanksANote{Correspondence: \texttt{chenwei@innocimda.com}},
\and
Zhaoke Huang\footnotemarkAAffil[2]$^{,}$
  \thanksANote{Correspondence: \texttt{zhaoke@innocimda.com}},
\and
Yiming Ma\thanksAAffil{Department of Biology, Hong Kong Baptist University, Hong Kong (SAR China)},
\and
Cunming Zhao\footnotemarkAAffil[3],
\and
Zhongying Zhao\footnotemarkAAffil[3],
\and
Hong Yan\footnotemarkAAffil[1]$^,$\footnotemarkAAffil[2]
}

\maketitle
\begin{abstract}

3D fluorescence microscopy is essential for understanding fundamental life processes through long-term live-cell imaging. However, due to inherent issues in imaging principles, it faces significant challenges including spatially varying noise and anisotropic resolution, where the axial resolution lags behind the lateral resolution up to 4.5 times. Meanwhile, laser power is kept low to maintain cell viability, leading to inaccessible low-noise and high-resolution paired ground truth (GT). 
To tackle these limitations, a dual Cycle-consistent Diffusion is proposed to effectively mine intra-volume imaging priors within 3D cell volumes in an unsupervised manner, $i.e.$, Volume Tells (VTCD), achieving de-noising and super-resolution (SR) simultaneously. Specifically, a spatially iso-distributed denoiser is designed to exploit the noise distribution consistency between adjacent low-noise and high-noise regions within the 3D cell volume, suppressing the spatially varying noise.
Then, in light of the structural consistency of the cell volume, a cross-plane global-propagation SR module propagates high-resolution details from the XY plane into adjacent regions in the XZ and YZ planes, progressively enhancing resolution across the entire 3D cell volume.
Experimental results on 10 $in$ $vivo$ cellular dataset demonstrate high improvements in both denoising and super-resolution, with axial resolution enhanced from $\sim 430~nm$ to $\sim 90~nm$.
\end{abstract}

\section{Introduction}
\label{sec:intro}

3D fluorescence confocal (FC) microscopy is crucial to reveal detailed volumetric structures within biological samples, which is often inaccessible in traditional 2D imaging.
When used for long-term live-cell observation, it provides the 3D context of complex biological processes, thereby making significant contributions to fields such as cell biology, neuroscience, and others by enabling the observation of these processes in their 3D context.
However, due to the mechanism of 3D FC imaging, it faces two major limitations, spatially varying noise and anisotropic resolution, as shown in Fig. \ref{FIRST_IMP}. 
Meanwhile, to maintain cell viability during long-term observation, the laser power must remain low, making it impossible to obtain high-resolution and low-noise 3D volumes, i.e., paired ground truth (GT) is inaccessible.
These limitations severely damage spatial structural information, hindering precise volumetric analysis in biology fields.

As an inherently ill-posed inverse problem, the image SR problem has benefited greatly from the rapid development of deep learning \cite{9044873}. Most of these methods are supervised, relying on paired HR-LR images, like regression-based deep learning, such as SRCNN \cite{7115171}, FSRCNN \cite{10.1007/978-3-319-46475-6_25} and denoising diffusion probabilistic models (DDPM) \cite{NEURIPS2020_4c5bcfec}. However, in 3D FC microscopy, the structural differences between individual cells during development make it impossible to obtain paired HR-LR images \cite{10.1007/978-3-031-43999-5_31,PMID:38039252}. 
Common unsupervised deep learning methods, such as learned degradation \cite{Bulat_2018_ECCV} (like CycleGAN \cite{Zhu_2017_ICCV}), Cycle-in-Cycle SR \cite{Yuan_2018_CVPR_Workshops} (like CinCGAN \cite{Yuan_2018_CVPR_Workshops}, CycleGAN \cite{Zhu_2017_ICCV}), and deep image prior \cite{Ulyanov_2018_CVPR}, fail to effectively exploit the unique imaging priors of 3D FC microscopy, leading to limit SR performance. Furthermore, the noise distribution in 3D FC microscopy varies spatially across different slices, making it challenging for existing deep learning methods to simultaneously perform de-noising and SR for 3D FC microscopy.

Current similar methods primarily focus on 3D FC microscopy under different imaging conditions, addressing either SR for low-resolution images/volumes or denoising for high-resolution images/volumes \cite{10.1093/gigascience/giad109,Hou2024,Shah:21}. For instance, Huang et al. \cite{Huang2023} used images from STEM as ground truth to enhance the resolution of confocal FC microscopy, while studies such as \cite{Ning2023, Qu2024, Park2022} employed self-supervised learning or Circle GAN to improve isotropic resolution restoration for volumetric FC microscopy. 
These methods either require more resource consumption to obtain high-resolution references or fail to exploit the internal priors of the 3D volume within a self-supervised learning framework.

To address the above limitations, this paper proposes a dual cycle-consistent diffusion (VTCD) for 3D FC microscopy, which leverages two distinct intra-volume imaging priors in 3D cell volumes to achieve simultaneous de-noising and SR during training, as described in the title, \emph{Volume Tells}. In the inference stage, the two generators in our trained model serve as the denoiser and the SR module, by inputting the low-resolution, high-noise 3D cell volume into the denoiser and SR module sequentially, a low-noise, high-resolution 3D cell volume can be obtained. Our method models the forward and reverse processes of de-noising and SR, enabling the model to accurately learn the intra-volume imaging priors in 3D cell volumes.
To summarize, our key contributions are four-fold:

$\bullet$ Our method exploits intra-volume imaging priors in 3D cell volumes as the forward stage of DDPM, and integrates DDPM into the cycle training, enabling simultaneous de-noising and SR for 3D FC microscopy.

$\bullet$ A spatially iso-distributed denoiser model the de-noising as a condition diffusion model, which learns the prior distribution of partial low-noise XY plane images and optimizes adjacent relatively high-noise regions, gradually reducing the spatially varying noise throughout the entire 3D cell volume without paired GT in the cycle training.

$\bullet$ A cross-plane global-propagation SR module enables the diffusion model to learn the prior distribution of continuous cell structures in the XY plane, propagating local structural information along the Z-axis within a more global space, continuously refining the super-resolved 3D cell volume without paired GT.

$\bullet$ Live-cell fluorescence datasets under diverse imaging conditions are collected, containing four-dimensional, single shot with great laser and axial imaging with high frequency. Extensive experimental results on these comprehensive datasets demonstrate significant improvements in SNR and resolution with multiple metrics.

\section{Related Works}

\subsection{Deep Learning-based SR}

Recent advances in deep learning have propelled the SR task to new heights, with a variety of architectures tailored to improve both image quality and computational efficiency.
Convolutional Neural Networks (CNNs) initially laid the foundation for SR with early models like SRCNN \cite{7115171}, VDSR \cite{Kim_2016_CVPR}, and EDSR \cite{Lim_2017_CVPR_Workshops}, which demonstrated the effectiveness of end-to-end learning for image SR.
Based on these, recent CNN-based SR methods have integrated attention mechanisms to focus on salient image regions, leading to marked improvements in SR performance and visual fidelity \cite{Zhang_2018_ECCV,8954252}.
The introduction of GANs has brought transformative advancements to SR, notably enhancing perceptual quality by enabling models to generate visually plausible details \cite{Ledig_2017_CVPR,Wang_2018_ECCV_Workshops,Wang_2021_ICCV}.
More recently, Transformer-based models have further revolutionized SR, leveraging self-attention mechanisms to capture long-range dependencies and improve representation capabilities \cite{Chen_2021_CVPR,Chen_2023_CVPR,NEURIPS2022_a37fea8e,Choi_2023_CVPR,Liang_2021_ICCV,Zhou_2023_ICCV}. IPT \cite{Chen_2021_CVPR} demonstrated the potential of Transformers pre-trained on large-scale datasets for SR. Subsequently, SwinIR \cite{Liang_2021_ICCV} introduced a hierarchical framework Swin Transformer \cite{Liu_2021_ICCV}, which efficiently processes images at multiple scales and has achieved state-of-the-art (SOTA) results.
Latest SR models include using graphs \cite{Tian_2024_CVPR}, diffusion model \cite{Zheng_2024_CVPR,Lee_2024_CVPR,Wang_2024_CVPR1}, dictionary learning\cite{Wang_2024_CVPR2}, but these methods either require a large amount of paired HR-LR data or their performance is not accurate and stable enough.

\subsection{SR of 3D Fluorescence Microscopy}

3D fluorescence microscopy SR has become a crucial research focus, aiming to address the inherent limitations of traditional fluorescence microscopy.
DFCAN \cite{Qiao2021} leveraged differences in frequency content across various features to learn hierarchical representations of high-frequency information related to diverse biological structures.
Park et al. \cite{Park2022} utilized a 3D optimal transport-driven CycleGAN network (OT-CycleGAN) for the isotropic recovery of various volumetric imaging data.
Self-Net \cite{Ning2023} utilized a self-learning strategy to enhance the axial resolution of the anisotropic raw data by employing high-resolution lateral images from the same dataset as target references.
Huang et al. \cite{Huang2023} designed a two-channel attention network (TCAN) to learn the mapping from low-resolution images to high-resolution ones.
Qu et al. \cite{Qu2024} proposed a Self-inspired Noise2Noise module (SN2N) that employs self-supervised data generation and a self-constrained learning process to enhance various super-resolution imaging modalities.
A method most comparable to this study is \cite{Qiao2024}, which utilized a SimCLR-like approach to simultaneously denoise and perform super-resolution on volumetric fluorescence microscopy. However, their method can't exploit the imaging prior of 3D cell volume, failing to achieve accurate and stable results. The comparison is presented in detail in experiments.

\section{Methodology}

\begin{figure*}
  \centering
  \includegraphics[width=0.95\linewidth]{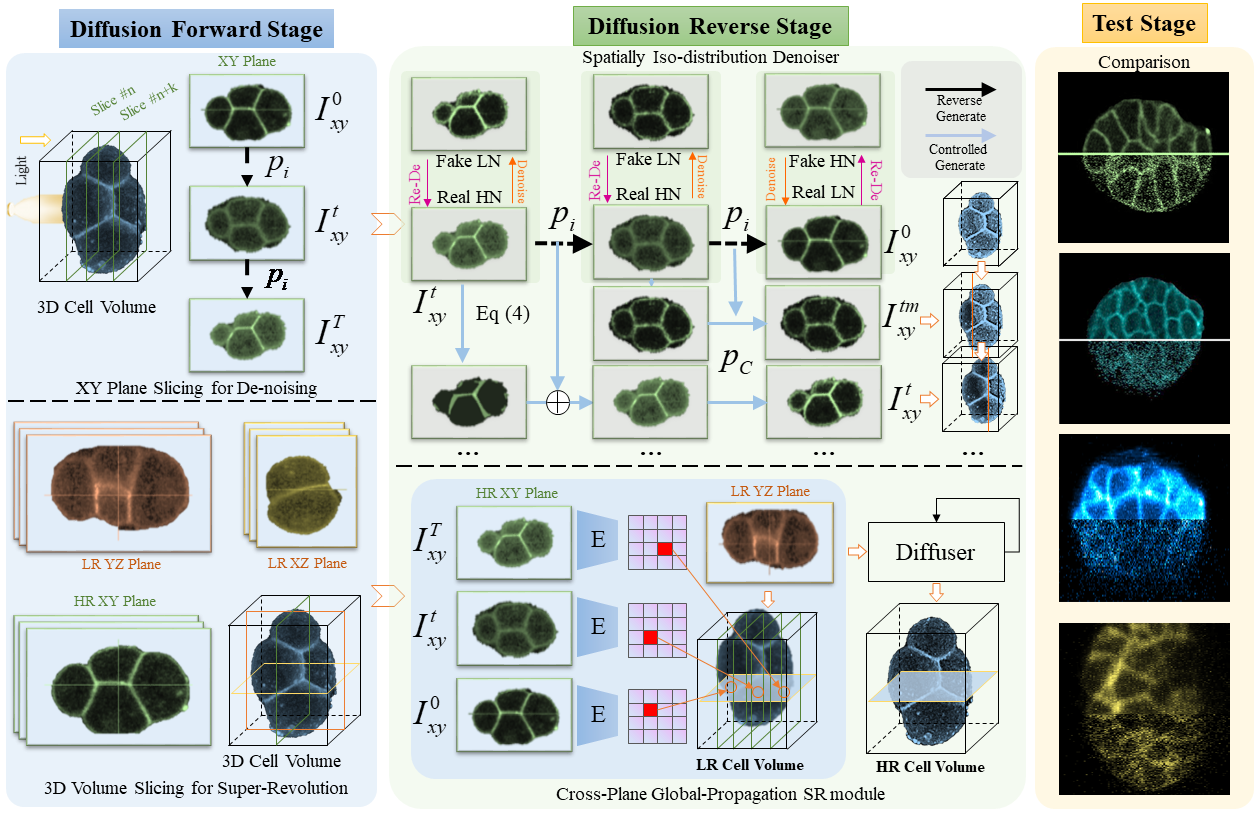}
  \caption{The overall framework of the proposed Method. \textbf{Left}: two targeted slicing strategies are modeled as the forward stage of diffusion model via the imaging prior for de-noising and SR 3D cell volume. \textbf{Middle}: the spatially iso-distributed denoiser model is a conditional diffusion model, which progressively reduces the noise of 3D cell volume along the Z-axis, and the cross-plane SRM enables the conditional diffusion model propagates the content distribution of the HR XY plane to the adjacent 3D volume space, eventually the whole 3D space. \textbf{Right}: The comparison of our results (above) and original slices (below). Compared to the original slice, our results show significantly reduced noise, and previously unobservable structural information becomes clear and discernible.}
  \label{fig_method}
\end{figure*}


\subsection{Overview of VTCD}

The image super-resolution (SR) problem, an inherently ill-posed inverse problem, has been significantly advanced by deep learning, though most methods rely on paired high- and low-resolution images, which are unavailable in 3D fluorescence microscopy due to structural differences between cells. Existing unsupervised methods, such as learned degradation and Cycle-in-Cycle SR, fail to leverage the unique priors of 3D fluorescence microscopy, limiting SR performance. 

As shown in Fig. \ref{fig_method}, since 3D FC microscopy is essentially an unsupervised problem, inspired by \cite{Park2022}, the VTCD is designed based on the cycle-in-cycle framework. For effectively mining intra-volume imaging priors, the VTCD will model the denoising and SR processes as two conditional diffusion denoising processes (CDDP). 

Given a 3D cell volume $V \in \mathbf{R}^{H\times W\times C}$, the $t$th XY/XZ/YZ plane slices $I_{xy}^t$, $I_{xz}^t$ and $I_{yz}^t$, where $H$, $W$ and $C$ is X-axis, Y-axis and Z-axis, $H$ and $W$ is high resolution, $Z$ is low resolution. The VTCD consists of two generators $G_r$ and $G_f$, and two discriminators $D_r$ and $D_f$, $r$ and $f$ notes the forward stage and reverse stage respectively. The spatially iso-distributed denoiser and cross-plane global-propagation SRM are noted as $SDD(\cdot)$ and $SRM(\cdot)$. 

In the forward stage, we utilize the characteristics of 3D FC microscopy to construct a special slicing method, which is equivalent to the forward stage of both CDDPs.
For the DDPM forward stage, the original diffusion model gradually adds Gaussian noise to the image $I_{0}$ at $t \in \{1, \dots, T\} $ to obtain the noisy image $I_{T}$.
For denoising in our case, in fluorescence microscopy, the light source is incident along the Z-axis, causing the XY plane slices along the Z-axis, $\{I_{xy}^0, I_{xy}^1, \dots, I_{xy}^T\}$ with increasing noise. Therefore, we perform continuous slicing along the Z-axis, obtaining a set of XY plane slices, as shown in the top-left part of Fig. \ref{fig_method}, which is equivalent to the forward stage, viewed as

\begin{equation}
     q(I_{xy}^t | I_{xy}^{t-1}) = N(\sqrt{1-\beta_t}I_{xy}^{t-1}; \beta_tI_d)
\end{equation}
Similar to the slicing above, for the forward stage of SR, the 3D volume slicing will produce three sets, XY/XZ/YZ plane slices, $\{I_{xy}^0, I_{xy}^1, \dots, I_{xy}^T\}$, $\{I_{xz}^0, I_{xz}^1, \dots, I_{xz}^T\}$, $\{I_{yz}^0, I_{yz}^1, \dots, I_{yz}^T\}$, as shown in the bottom-left part of Fig. \ref{fig_method}, can be viewed as
\begin{equation}
     q(I_{tmp}^t | I_{xy}^{t-1}) = N(\sqrt{1-\beta_t}I_{xy}^{t-1}; \beta_tI_d)
\end{equation}
where $I_{tmp}^t$ is the $t$th slice of XZ/YZ planes.

In the reverse stage of the denoising CDDP, the spatially iso-distributed denoiser guides the generative trajectory via the semantic content of the corresponding slice. 
In the reverse stage of the SR CDDP, the cross-plane global-propagation SRM maps the XY-plane slice features to the corresponding coordinates of a generated 3D grid, guiding the SR generative trajectory.
Next, the spatially iso-distributed denoiser and the cross-plane global-propagation SR module will be introduced in detail with the description of the reverse stage.

\subsection{Spatially Iso-Distributed Denoiser}

In the reverse stage of denoising CDDP, the key is to control the generative trajectory to obtain the LN $\hat{I}_{xy}^{t}$ rather $I_{xy}^{0}$. Thus, the spatially iso-distributed denoiser uses the semantic content of HN slice $I_{xy}^{t}$ to guide the generative trajectory at the reverse stage. 


In the reverse stage, the original diffusion model denoises a latent sample $\hat{I}_{xy}^{T}$ to a data sample $I_{xy}^{0}$ as 
\begin{equation}
     I_{xy}^{t-1} = \frac{1}{\sqrt{1-\beta_t}}(I_{xy}^t - \frac{\beta_t}{\sqrt{1-\alpha_t}}; \epsilon_t^\theta(I_{xy}^t))+\sigma_t z_t
\end{equation}

However, we aim to obtain the low noise $T$th slice $I_{xy}^{T}$ rather than $I_{xy}^{0}$, the original DDPM is unable to address this issue. Thus, We guide the generative trajectory in the reverse process of $I_{xy}^t$ using the shape and semantic information of $I_{xy}^t$, as shown in Fig. \ref{fig_method}. Following \cite{shen2020interpreting}, We define the semantic consistency of the latent encoding with respect to the semantic information of $I_{xy}^t$ as:

\begin{equation}
d(I_{xy}^t, n) = n^{T} I_{xy}^t
\end{equation}
where $n$ is the unit normal vector of the hyperplane. Then the latent encodings can be edited by $x'_t = x_t + \lambda d(x_t, s_t)$. At the same time, we use the contextural loss to better control the generated results, which will be described in detail in supplementary materials. The reverse stage can be presented as 
\begin{equation}
     x'_{t-1} = \frac{1}{\sqrt{1-\beta_t}}(x'_t - \frac{\beta_t}{\sqrt{1-\alpha_t}}; \epsilon_t^\theta(x'_t))+\sigma_t z_t
\label{eq_updata}
\end{equation}

The distribution prior progressively de-noises XY plane slices along the Z-axis.
Through continuously editing this reverse stage at every slice of $\{I_{xy}^0, I_{xy}^1, \dots, I_{xy}^T\}$, the 3D cell volume can be denoised with the imaging prior, and enhance the semantic information of every slice to preserve its true cell shape. 
This progressive denoising effectively reduces the spatially varying noise across the entire 3D cell volume.

\subsection{Cross-Plane Global-Propagation SR module}

For the reverse stage of SR's CDDP, in our case, the high resolution of the XY plane serves as a guide for the SR of the XZ and YZ planes, ultimately resulting in an HR 3D cell volume. However, most existing diffusion models rely on ample HR 3D data to supervise this process and get the 3D cell volume. In our case, such ample HR 3D cell volumes are evidently unavailable, leaving only the 2D HR slices of the XY plane as a resource.

Thus, in the reverse stage of SR, the cross-plane global-propagation SRM is designed to propagate the HR information from the XY plane to the surrounding 3D cell volume. This module generates intermediate 3D-aware features, then progressively refines and reconstructs the 3D cell volume, ensuring that the anisotropic resolution is restored while maintaining consistency with the high-resolution structural details from the XY plane.

Given a 3D cell volume $ V \in \mathbb{R}^{H \times W \times C} $, where $ H $, $ W $, and $ C $ represent the X, Y, and Z axes, respectively, with $ H $ and $ W $ corresponding to high-resolution dimensions and $ C $ corresponding to a low-resolution dimension, we extract three slices of the volume: the XY, XZ, and YZ plane slices, denoted as $ I_{xy}^t $, $ I_{xz}^t $, and $ I_{yz}^t $. Following the approach outlined in \cite{henzler2021unsupervised, karnewar2023holodiffusion}, we first project the 3D coordinates $ x_{hwc}^V $ of each grid element $ (h, w, c) $ from the 3D cell volume onto each corresponding XY plane slice $ I_{xy} $. A fixed ResNet32 is then applied to extract the 2D features $ f_{hwc}^t $ for each slice, with $ t $ indicating the specific XY plane slice.

The result is a generated grid $ \bar{V} \in \mathbb{R}^{d \times H \times W \times C} $, where each grid element is associated with a corresponding 2D feature. To enhance the resolution of $ \bar{V} $, an Accumulator MLP, $ A_{acc} $, is employed to aggregate the surrounding spatial information and produce a new grid element. Specifically, $ A_{acc} $ takes as input the features $ f_{mno} $ from neighboring grid elements, where $ m = h \pm \text{shift} $, $ n = w \pm \text{shift} $, and $ o = c \pm \text{shift} $, with $ \text{shift} $ set to 1 in our implementation. The output of $ A_{acc} $ is a feature vector $ \theta \in \mathbb{R}^k $, where $ k $ corresponds to the total number of possible shifts. This process can be interpreted as

\begin{equation}
    f_{hwc} = \sum_j^k \theta_j f_{mno}^j
\end{equation}
Thus, an HR grid $\bar{V} \in \mathbb{R}^{d \times H \times W \times C}$ is generated by cross-plane propagating the HR information of XY plane. In the reverse stage of SR, we only need to update $I_{xz}^t$ and $I_{yz}^t$ from XZ/YZ plane by 

\begin{equation}
    {I'}_{xz}^t = I_{xz}^t + SL \left( V \right)
\end{equation}
where $SL(\cdot)$ represents the operation that forms a slice from the features of $ I_{xz}^t $ at the corresponding spatial location $\bar{V}$, and overlays it onto the original LR slice. Subsequently, Eq.~\ref{eq_updata} is applied in the reverse stage.

The SR process leverages the structural consistency of the 3D cell volume to propagate HR information from the XY plane to adjacent spatial regions along the Z-axis. Through the step-by-step application of the aforementioned process to XZ/YZ plane slices, the high resolution of the XY plane is globally propagated throughout the 3D volume. 
Thus, the SR process ensures that HR information from the XY plane is effectively extended to the entire 3D cell volume, achieving anisotropic super-resolution with structural consistency.

Due to space limitations, the training strategy and loss are included in the supplementary materials.

\section{Experiments}

\subsection{The Novel Dataset and Experimental Setup}

\begin{table}[htbp]
  \centering
    \begin{tabular}{@{}lclc@{}}
    \toprule
    Dataset & Image Num. & Axial Res. & Laser Intens.\\
    \midrule
    UnpairedTrain & 22560 & 94 & Middle\\
    FullrefXY-1 & 1028 & 94 & Low\\
    FullrefXY-2 & 584 & 92 & Middle\\
    FullrefXY-3 & 456 & 30 & High\\
    NorefZ-1 & 887 & 94 & Middle \\
    NorefZ-2 & 1098 & 200 & High \\
    NorefZ-3 & 487 & 30 & High \\
    \bottomrule
  \end{tabular}
  \caption{The training and benchmark evaluation dataset we provided for all FC image processing (cell analyses especially).}
  \label{table1}
\end{table}

\begin{figure*}[htbp]
  \centering
  \begin{subfigure}{1\linewidth}
  \centering
    \includegraphics[width=0.90\linewidth]{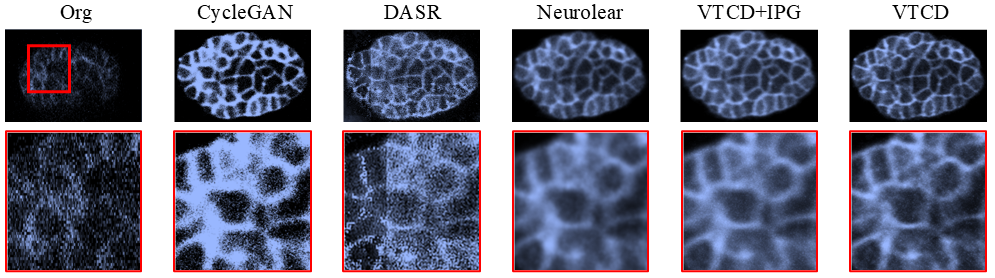}
    \label{performance-a}
    \caption{Comparison of results in the XY plane near the upper Z-axis region, showing differences in detail preservation and clarity.}
  \end{subfigure}
  
  \begin{subfigure}{1\linewidth}
  \centering
    \includegraphics[width=0.90\linewidth]{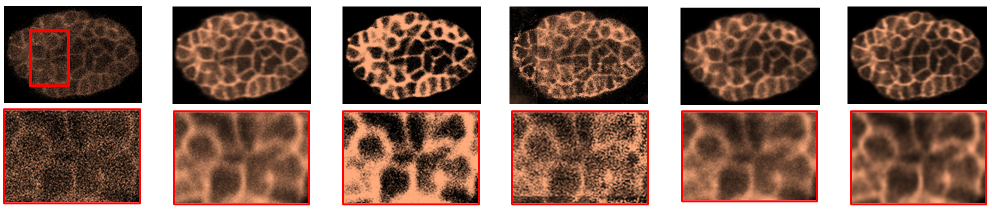}
    \label{performance-b}
    \caption{Comparison of results in the XZ plane, highlighting differences in structure and resolution.}
  \end{subfigure}
  \begin{subfigure}{1\linewidth}
  \centering
    \includegraphics[width=0.90\linewidth]{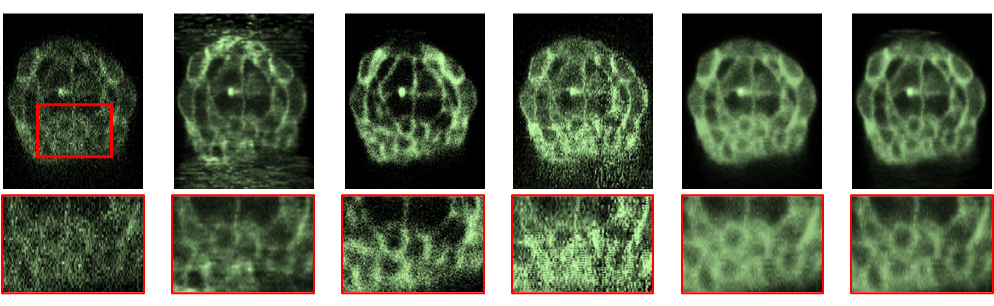}
    \label{performance-c}
    \caption{Comparison of results in the YZ plane, illustrating differences in structure and resolution.}
  \end{subfigure}
  \caption{The qualitative comparison of the cell fluorescence dataset from three extremely different aspects. These results are from different cell and different life time, thus there are different number and shape of cells.}
  \label{performance1}
\end{figure*}

All imaging datasets were collected from live-cell embryos, using either the Leica SP5 II or Stellaris confocal microscopy systems. The images were subsequently processed and analyzed with Leica LAS X software and ImageJ. For 4D data acquisition, to maintain embryo viability, light intensity was kept low by using a scanning speed of 8000 Hz with a water immersion objective, which resulted in significant degradation, noise, and spatially anisotropic signals. High-frequency axial and single-shot images were captured with a higher laser power for observation purposes only, not for data curation. The imaging was divided into time blocks, with each segment corresponding to a 60-time-point interval. The z-axis compensation ranged from 0.5\% to 3\% for the 488 nm laser and from 20\% to 95\% for the 594 nm laser.

As demonstrated in Table \ref{table1}, we provide a total of 600 new 3D fluorescence cell volumes (22560 lateral 2D images, UnpairedTrain) for training and evaluation which was not available before. With higher weights for late cell stage embryos, we randomly generate unpaired ~11000 images as a training dataset. To construct a benchmark for cell fluorescence imaging super-resolution application, from these 2D images, we generate 3 evaluation datasets with full reference (FullrefXY-1, FullrefXY-2, FullrefXY-3) and other 3 with no reference (NorefZ-1, NorefZ-2, NorefZ-3) for a comprehensive comparison between our method and existing SOTA unsupervised learning methods. Full reference datasets are constructed on XY plane images, while the input images are downscale 4 times (128$\times$178) and the reference (evaluating GT) is the original XY images (512$\times$712). 

The model training used mCherry-labeled (cell membrane stained) fluorescence images as inputs, without the HR
GT images serve as the learning targets. The real 4D volumes are in the shape of 512$\times$712$\times$94 for keeping cells healthy. The training was optimized using the Adam optimizer, starting with an initial learning rate of \(5 \times 10^{-3}\) and a weight decay of \(1 \times 10^{-5}\), along with the AMSGrad variant for improved gradient descent. The proposed model and other methods are all trained on 20 CPUs (Intel(R) Xeon(R) Platinum 8457C), and 3 NVIDIA A100 GPUs.

\subsection{Qualitative De-noising and Super-resolution Comparisons}
To further evaluate the effects of our method, we compared VTCD and VTCD-based methods with other existing unsupervised learning models \cite{Park2022,Yuan_2018_CVPR_Workshops,Zhu_2017_ICCV,Chen_2023_CVPR,Tian_2024_CVPR,wang2021unsupervised}. On the same cell membrane fluorescence GT (Table \ref{table1}, unpaired images in low and high resolution respectively). Comprehensive qualitative comparisons demonstrate the significant improvement from VTCD in Fig. \ref{performance1} and Supplementary Information. The de-noising effects perform on all XY, YZ, and XZ planes while the super-resolution module solves the isotropic low-resolution issues in the YZ and XZ planes. It is obvious that in the raw fluorescence images, the cell membrane signals are almost impossible to recognize in the Z-direction (Fig. \ref{performance1}). These optical and physical restrictions hinder the resolution of the third axis Z. We demonstrate that our method VTCD dramatically interpolates the information at the YZ and XZ plane and makes the Z-direction viewing clear enough for human eye recognition (Supplementary Note). 

Besides the great enhancements from VTCD for the raw images, comparing other existing good unsupervised learning methods, our proposed method has outperformed them qualitatively. No matter from local texture restoration, global noise elimination, or Z-axis super-resolution interpolation, VTCD has the best interpretation for the cell structural signals. For example, unlike the chaotic and global noise in CycleGAN and Neuroclear \cite{Zhu_2017_ICCV,Park2022} second and fourth columns in Fig. S1, the results from VTCD are much cleaner. In DSAR's results, the super-resolution for YZ and XZ eliminates some signals when the cell densities are high (at the left part of the embryos) but VTCD could balance these issues in a good way for detailed observations.


\subsection{Evaluation on Dataset with Full and No Reference Metrics}

\begin{table*}[htbp]
    \centering
    \begin{tabular}{m{1.8cm}|m{2cm}|m{1.2cm}|m{1.2cm}|m{1.2cm}|m{1.2cm}|m{1.2cm}|m{1.2cm}|m{1.2cm}|m{1.2cm}}
        \toprule \toprule
        \multirow{2}{*}{\textbf{Dataset}} & \multirow{2}{*}{\textbf{Method}} &  \multirow{2}{*}{\textbf{Years}} & \multicolumn{4}{c|}{\textbf{Full Reference Metrics}} & \multicolumn{3}{c}{\textbf{No Reference Metrics}} \\ 
        \cline{4-10}
        ~ & ~ & ~ & PSNR$\uparrow$ & SSIM$\uparrow$& LPIPS$\downarrow$ & VIF$\uparrow$ & NIQE$\downarrow$ & PIQE$\downarrow$ & NRQM$\uparrow$ \\ \midrule

        \multirow{7}{*}{\textbf{FullrefXY-1}} 
            & CycleGAN & 2018 & 32.5422 & 0.6200 & 0.4576 & 0.0533 & 21.2918 & 47.3730 & 4.4334 \\ 
            & CinCGAN & 2019 & 31.1730 & 0.5997 & 0.3204 & 0.0537 & 20.9956 & 42.2347 & 4.0328 \\ 
            & DASR & 2021 & 32.2968 & 0.6197 & 0.4876 & 0.0376 & 20.1543 & 45.1351 & 4.5786 \\ 
            & Neuroclear & 2022 & 30.5578 & 0.5798 & 0.3122 & 0.0487 & 21.8716 & 48.1651 & 4.4882 \\  
            & \textbf{VTCD+HAT} &  2023 & 32.3669 & 0.6236 & {\color{blue} \textbf{0.3069}} & 0.0363 & {\color{blue} \textbf{16.8470}} & 49.9810 & 4.3840 \\ 
            & \textbf{VTCD+IPG} &  2024 & {\color{blue} \textbf{32.5718}} & {\color{blue} \textbf{0.6338}} & {\color{red} \textbf{0.3040}} & {\color{blue} \textbf{0.0630}} & {\color{red} \textbf{16.9656}} & {\color{blue} \textbf{41.5012}} & {\color{red} \textbf{5.0619}} \\ 
            & \textbf{VTCD} &  Ours & {\color{red} \textbf{32.5719}} & {\color{red} \textbf{0.6357}} & 0.3221 & {\color{red} \textbf{0.0653}} & 17.0562 & {\color{red} \textbf{41.0117}} & {\color{blue} \textbf{4.6480}} \\ \midrule

        \multirow{7}{*}{\textbf{FullrefXY-2}} 
            & CycleGAN & 2018 & 38.5692 & 0.9766 & 0.7076 & 0.0372 & 21.6533 & 48.2666 & 4.3256 \\ 
            & CinCGAN & 2019 & 35.3893 & 0.5896 & 0.5311 & 0.0297 & 20.8319 & 52.3552 & 3.5771 \\ 
            & DASR & 2021 & 38.5214 & 0.9612 & 0.4682& 0.0430  & 20.4517 & 47.2133 & 3.5114 \\ 
            & Neuroclear & 2022 & 36.7488 & 0.8405  & 0.5654 & 0.0281& 19.5124 & 52.9998 & 3.5871 \\ 
            & \textbf{VTCD+HAT} & 2023 & {\color{blue} \textbf{39.8612}} & {\color{blue} \textbf{0.9826}}  & {\color{blue} \textbf{0.3648}} & 0.0369 & 20.5301 & 48.6379 & 4.5130 \\ 
            & \textbf{VTCD+IPG} &2024  & 37.9154 & 0.9519 & {\color{red} \textbf{0.3138}} & {\color{red} \textbf{0.0700}} & {\color{blue} \textbf{18.9704}} & {\color{blue} \textbf{46.4613}} & {\color{red} \textbf{4.6895}} \\ 
            & \textbf{VTCD} & Ours & {\color{red} \textbf{40.7851}} & {\color{red} \textbf{0.9866}} & 0.3708 &  {\color{blue} \textbf{0.0566}} &{\color{red} \textbf{15.6277}} & {\color{red} \textbf{44.7752}} & {\color{blue} \textbf{4.5877}} \\ \midrule

        \multirow{7}{*}{\textbf{FullrefXY-3}} 
            & CycleGAN & 2018 & 33.6425 & 0.7860 & 0.2557 & 0.0516 & 20.9541 & 54.5335 & 4.0513 \\ 
            & CinCGAN & 2019 & 33.4772 & 0.5198 &  0.3766 & 0.0731 & 19.4749 & 53.5507 & 3.6956 \\ 
            & DASR & 2021 & 34.2431 & 0.8066 & 0.2777 & 0.0234 & 16.8766 & 50.2188 & 3.8611 \\ 
            & Neuroclear & 2022 & 33.5124 & 0.7339 & 0.3665 & 0.0486 & 16.2343 & 53.9991 & 3.8761 \\ 
            & \textbf{VTCD+HAT} & 2023  & 34.1687 & 0.8053 & 0.2077 & {\color{blue} \textbf{0.0989}} & 16.0564 & {\color{blue} \textbf{48.9741}} & 4.0874 \\ 
            & \textbf{VTCD+IPG} & 2024 & {\color{blue} \textbf{34.8173}} & {\color{blue} \textbf{0.8081}} & {\color{blue} \textbf{0.1995}} & 0.0954 & {\color{red} \textbf{15.3609}} & {\color{red} \textbf{48.6838}} & {\color{red} \textbf{4.4972}} \\ 
            & \textbf{VTCD} & Ours & {\color{red} \textbf{34.9251}} & {\color{red} \textbf{0.8137}} & {\color{red} \textbf{0.1986}} & {\color{red} \textbf{0.1106}} & {\color{blue} \textbf{15.9362}} & 53.4817 & {\color{blue} \textbf{4.2382}} \\ \bottomrule \bottomrule 

    \end{tabular}
    \caption{The comparison results on 3 full reference datasets. Multiple metrics are applied in the experiments for fair and comprehensive evaluation. $\uparrow$ means higher value is better performance; $\downarrow$ indicates that smaller value is better.}

    \label{table2}
\end{table*}


The quantitative evaluation of our VTCD model was conducted against 6 SOTA unsupervised learning methods across both full-reference and no-reference metrics on various datasets, including FullrefXY-1, FullrefXY-2, FullrefXY-3, and NorefZ-1 to NorefZ-3. The results consistently demonstrate the superior performance of our proposed VTCD in FC images' de-noising and super-resolution.

\begin{figure}[t]
  \centering
   \includegraphics[width=0.95\linewidth]{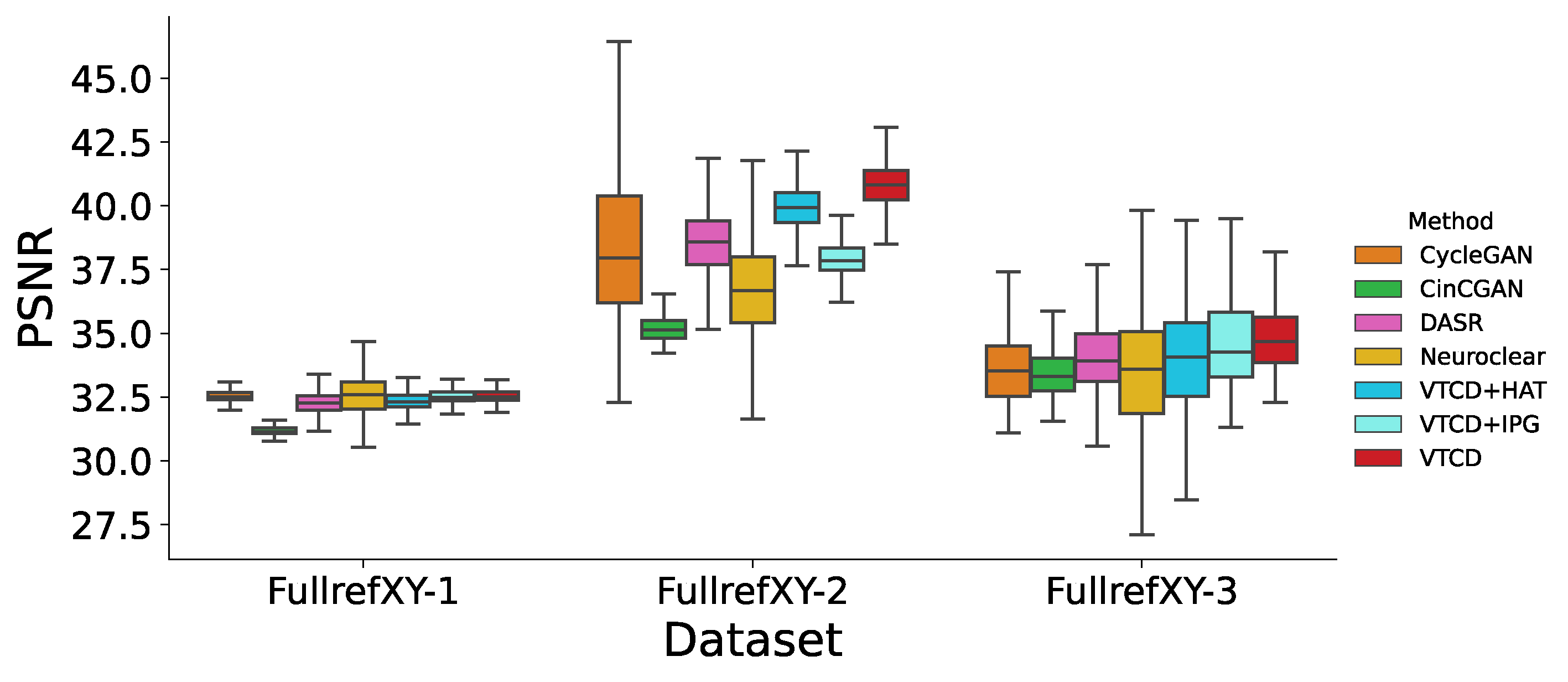}

   \caption{Comparison of PSNR values across different methods in 3 datasets (under multiple imaging conditions), highlighting the statistical performance variations.}
   \label{fig4}
\end{figure}

For the FullrefXY-1, FullrefXY-2, and FullrefXY-3 datasets, our VTCD model consistently outperforms competing methods in both PSNR and SSIM metrics. In FullrefXY-1, VTCD achieves a PSNR of 32.5719 ($\sim$6.59\% over Neuroclear) and an SSIM of 0.6357 ($\sim$2.6\% over CycleGAN), surpassing prior models such as CinCGAN \cite{Yuan_2018_CVPR_Workshops} and DASR \cite{wang2021unsupervised}. The model also performs competitively across perceptual metrics, maintaining an LPIPS score of 0.3221 and achieving a lower PIQE of 41.0117 compared to other baseline models, indicating a significant reduction in perceived image distortions. In FullrefXY-2, VTCD attains a PSNR of 40.7851 ($\sim$5.9\% over DASR) and an SSIM of 0.9866 with the best. VTCD also excels in reducing perceptual artifacts, as evidenced by its NIQE score of 15.6277, which is notably lower than that of most competitors, showcasing its robustness in preserving image details and clarity in super-resolution tasks. The model's performance is further validated by its NRQM score of 4.5877, highlighting a well-balanced approach to enhancing perceptual quality. These metrics reflect VTCD's effective handling of both image fidelity and perceptual quality (Fig. \ref{fig4}).

In the axial super-resolution tasks, where ground truth is unavailable, the NorefZ-1, NorefZ-2, and NorefZ-3 datasets were used for evaluation using no-reference metrics. VTCD demonstrates notable strengths, particularly in PIQE and NRQM scores, which are indicators of image quality and perceptual realism. In NorefZ-1, VTCD achieves a PIQE of 34.7048, reflecting a 9.8\% improvement over VTCD+IPG \cite{Tian_2024_CVPR}, and maintains a competitive NIQE score of 15.9237. Similarly, in NorefZ-2, VTCD attains the highest NRQM of 5.7390, emphasizing the model's ability to achieve reduced noise and maintain image quality across challenging no-reference scenarios.

Overall, these quantitative evaluation results confidently demonstrate the superior performance of VTCD across both full-reference and no-reference metrics. The model not only excels in enhancing image clarity and reducing perceptual artifacts but also maintains robustness in preserving fine details, demonstrating its effectiveness across diverse datasets.

\begin{table}[t]
    \centering
    \begin{tabular}{m{1.3cm}|m{1.9cm}|m{1.0cm}|m{1.0cm}|m{1.1cm}}
        \toprule \toprule
        \multirow{2}{*}{\textbf{Dataset}} & \multirow{2}{*}{\textbf{Method}} & \multicolumn{3}{c}{\textbf{No Reference Metrics}} \\ \cline{3-5}
                                    & & NIQE$\downarrow$ & PIQE$\downarrow$ & NRQM$\uparrow$ \\ \hline

        \multirow{7}{*}{\textbf{NorefZ-1}} 
            & CycleGAN & 21.1589 & 45.6363 & 3.5341 \\ 
            & CinCGAN & 16.4999 & 49.9127 & 2.9472 \\ 
            & DASR  & 16.9833 & 49.5444 & 3.5225 \\ 
            & Neuroclear  & 16.8188 & 48.0947 & 3.7322 \\ 
            & \textbf{VTCD+HAT}  & 16.1298 & 43.9811 & {\color{blue} \textbf{4.3409}} \\ 
            & \textbf{VTCD+IPG}  & {\color{red} \textbf{15.7458}} & {\color{blue} \textbf{38.5549}} & {\color{red} \textbf{5.1336}} \\ 
            & \textbf{VTCD}  & {\color{blue} \textbf{15.9237}} & {\color{red} \textbf{34.7048}} & 4.0482 \\ \midrule

        \multirow{7}{*}{\textbf{NorefZ-2}} 
            & CycleGAN  & 16.2938 & 45.3978 & 4.3399 \\ 
            & CinCGAN  & 16.8824 & 49.5337 & 4.8908 \\ 
            & DASR  & 14.5744 & 47.6895 & 5.3026 \\ 
            & Neuroclear  & 16.3213 & 48.1098 & 4.9018 \\ 
            & \textbf{VTCD+HAT}  & 13.9311 & {\color{blue} \textbf{46.5909}} & 5.2468 \\ 
            &\textbf{VTCD+IPG}  & {\color{red} \textbf{12.8717}} & 46.6903 & {\color{blue} \textbf{5.6680}} \\ 
            & \textbf{VTCD}  & {\color{blue} \textbf{13.4206}} & {\color{red} \textbf{46.5737}} & {\color{red} \textbf{5.7390}} \\ \midrule

        \multirow{7}{*}{\textbf{NorefZ-3}} 
            & CycleGAN & 18.3970 & 66.9309 & 2.4942 \\ 
            & CinCGAN  & 14.8976 & 73.5437 & 2.3943 \\ 
            & DASR  & 12.8424 & 76.5410 & 3.6288 \\ 
            & Neuroclear  & 19.3217 & 69.1047 & 2.2472 \\ 
            & \textbf{VTCD+HAT}   & {\color{blue} \textbf{12.2402}} & {\color{blue} \textbf{60.2010}} & {\color{blue} \textbf{5.7724}} \\ 
            & \textbf{VTCD+IPG}  & {\color{red} \textbf{12.1711}} & 62.7601 & 4.8728 \\ 
            & \textbf{VTCD}   & 12.4983 & {\color{red} \textbf{53.2591}} & {\color{red} \textbf{6.3571}} \\ \bottomrule \bottomrule

    \end{tabular}
    \caption{The comparison results on 3 no reference datasets. Three no reference metrics (DNN pre-trained model) are used in calculating the performance of the super-resolution images. $\uparrow$ means higher value is better performance; $\downarrow$ indicates that smaller value is better.}

    \label{table3}
\end{table}

\subsection{Ablation Studies}

\begin{table}[h!]
  \centering
    \begin{tabular}{m{2.0cm}|m{0.6cm}|m{0.6cm}|m{0.6cm}|m{1.1cm}|m{1.1cm}}
    \toprule
    Module Name & S-D & C-S & DcL & PSNR & SSIM \\
    \midrule
    Base Model  & {\color{gray} \ding{55}} & {\color{gray} \ding{55}} & {\color{gray} \ding{55}} & 31.9431 & 0.6211 \\
    +SD  & $\checkmark$ & {\color{gray} \ding{55}} & {\color{gray} \ding{55}} & 32.3147 & 0.6298 \\
    +SD+CS & $\checkmark$ & $\checkmark$ & {\color{gray} \ding{55}} & {\color{blue} \textbf{32.4987}} & {\color{blue} \textbf{0.6308}} \\
    +SD+DcL & $\checkmark$ & {\color{gray} \ding{55}} & $\checkmark$ & 32.4237 & 0.6278 \\
    +SD+CS+DcL & $\checkmark$ & $\checkmark$ & $\checkmark$ & {\color{red} \textbf{32.5719}} & {\color{red} \textbf{0.6357}} \\
    \bottomrule
  \end{tabular}
  \caption{The ablation studies on 2 proposed modules and novel loss functions with FullrefXY-1 dataset. SD denotes SID-Denoiser, CS denotes  CPGP-SRM, and DcL denotes Dual Cycle-consistent Losses.}
  \label{table4}
\end{table}

Without SID-Denoiser: When the SID-Denoiser was removed, we observed a significant reduction in the de-noising capability. The PSNR dropped by approximately 1.16\%, and SSIM decreased from 0.6298 to 0.6211, indicating that the model struggled to suppress spatially varying noise effectively. This highlighted the importance of spatially distributed noise handling in achieving high-quality 3D reconstructions.

Without CPGP-SRM: Removing the CPGP-SRM led to a decline in super-resolution quality, especially along the z-axis. The PSNR decreased by 0.57\%, and the visual inspection showed a lack of clarity in the XZ and YZ planes. This confirmed the necessity of global-propagation to enhance resolution anisotropically and propagate XY-plane high-resolution details.

Without Dual Cycle-consistent Losses: The exclusion of cycle-consistent loss resulted in both reduced PSNR and SSIM, suggesting less effective feature consistency across different planes. The model's PSNR dropped by 0.46\%, and the resulting images showed visible structural inconsistencies, particularly along the boundaries of cellular structures.

\section{Conclusion}

The proposed VTCD leverages the unique imaging priors of 3D fluorescence microscopy, and within the framework of cycle training, it models de-noising and SR as two conditional diffusion models without paired GT. This approach addresses the limitations of current deep learning-based methods in de-noising and SR for 3D fluorescence microscopy, while simultaneously achieving de-noising and SR of 3D cell volumes. 
Through the dual objectives of de-noising and SR, VTCD not only restores the suppressed details in the Z-axis but also enhances the structural consistency across 3D cell volumes. This dual capability is particularly beneficial for understanding fundamental life processes in long-term time-lapse live-cell studies. Moreover, the VTCD framework is designed to generalize across various experimental conditions, making it a versatile tool for improving the quality of 3D fluorescence microscopy data.

{\small
\bibliographystyle{utils/ieee_fullname}
\bibliography{bibs/egbib}
}

\end{document}


\title{Supplementary of Paper $-$ Volume Tells: Dual Cycle-Consistent Diffusion for 3D Fluorescence Microscopy De-noising and Super-Resolution}


\author{Zelin Li\\
Institution1\\
Institution1 address\\
{\tt\small firstauthor@i1.org}
\and
Zhaoke Huang\\
Institution2\\
First line of institution2 address\\
{\tt\small secondauthor@i2.org}
}

\maketitle

\section{The Motivation in Details}

Fluorescence microscopy (FM) is a powerful tool for studying cellular processes by using fluorescent markers to label components like membranes, nuclei, and proteins, enabling detailed visualization of cell structures, functions, and dynamics. It provides insights into processes such as cell signaling, gene expression, and tissue organization while revealing critical features like shape, volume, and subcellular localization. By tracking these features over time, FM uncovers dynamic processes such as cell division, migration, and apoptosis, as well as cellular heterogeneity. However, achieving continuous, high-quality signals for precise observation and analysis remains a challenge, especially in complex, multi-cellular contexts where rapid divisions and small cell sizes blur fluorescence signals and reduce resolution.

The limitations in 3D fluorescence imaging primarily stem from the inherent characteristics of the imaging mechanism. First, due to the optical diffraction limit, the system achieves higher resolution in the lateral (XY) plane but has lower resolution along the axial (Z) axis. This anisotropic resolution difference is caused by the physical properties of the optical system: the focusing ability is weaker along the axial direction, making it challenging to capture the same level of detail in depth as in the lateral plane. Additionally, the uneven noise distribution is related to how light propagates through the sample. 
In deeper regions along the Z-axis, light undergoes more absorption and scattering, causing signal attenuation and resulting in a lower signal-to-noise ratio (SNR), which in turn increases noise in these deeper layers. 
Variations in tissue density and optical thickness further impact light penetration, leading to higher signal quality near the sample surface and more interference in deeper layers. 
These optical limitations affect the overall quality of 3D fluorescence imaging, particularly in deeper structures, where they lead to detail loss and increased noise. 
What's more, ground truth (GT) is hard to obtain. Without GT data for deep structures, accurately resolving these regions becomes even more challenging, as there is no reference for training models to reconstruct high-fidelity images. 
These optical limitations, combined with the lack of GT, affect the overall quality of 3D fluorescence imaging, particularly in deeper structures, leading to detail loss and increased noise.

Recent advances in deep learning have demonstrated the potential of generative models to enhance FM imaging, with diffusion models emerging as particularly effective for both denoising and super-resolution tasks, as described in the section of related work. Diffusion models are probabilistic generative approaches that learn to reverse a stochastic noise addition process, allowing them to iteratively refine noisy images. This is especially well-suited for FM imaging, where noise is pervasive, as these models can systematically model and remove noise while preserving biologically relevant details.

Our work contributes to this growing field by developing a algorithm that facilitates the ambiguous and low-resolution fluorescence images clear and visible. VTCD has been applied to study a variety of biological systems, providing insights into factors like cell signaling pathways, intercellular interactions, and the impact of genetic or environmental perturbations.

\section{Additional Experimental Demonstrations}

\subsection{The 2D Quantitative Demonstration}

\begin{figure*}[t]
  \centering
   \includegraphics[width=0.90\linewidth]{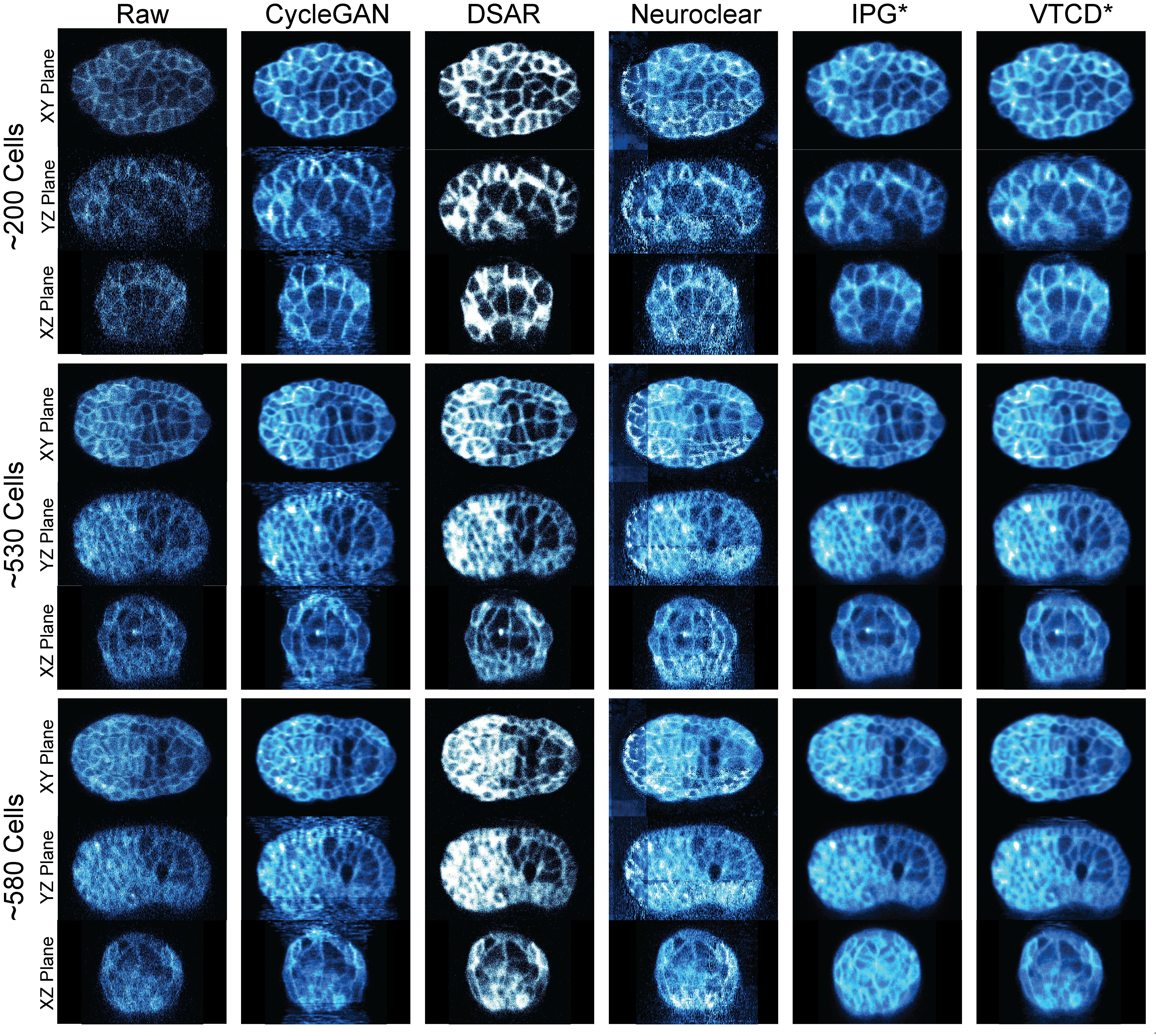}

   \caption{The qualitative comparisons on fluorescence images between multiple methods (de-noising and super-resolution).}
   \label{figS1}
\end{figure*}

\begin{figure*}[t]
  \centering
   \includegraphics[width=0.85\linewidth]{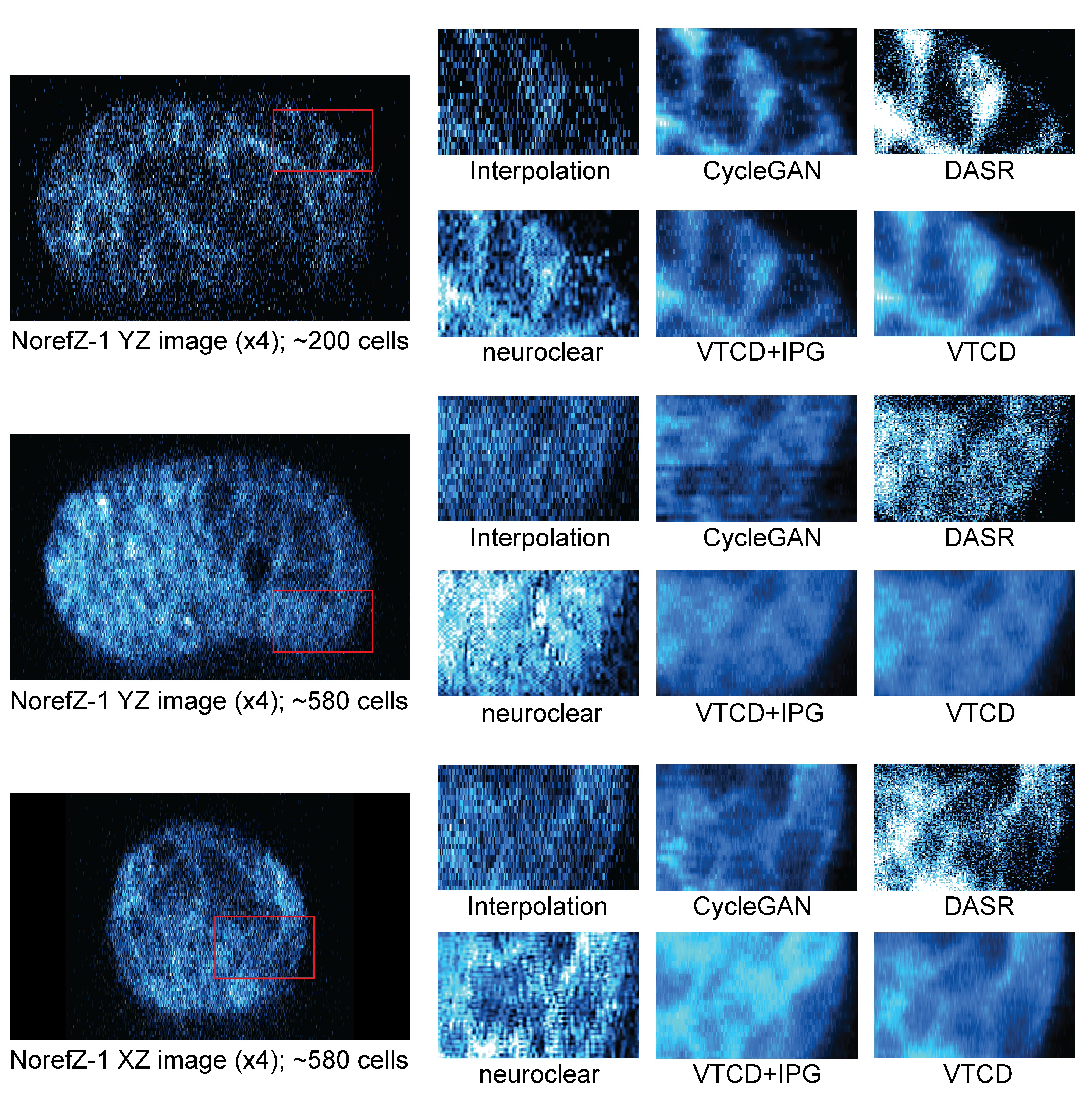}

   \caption{In details, the qualitative comparisons on fluorescence images between multiple methods (de-noising and super-resolution).}
   \label{figS2}
\end{figure*}
To further validate the effectiveness of our cycle-consistent diffusion architecture, we show more details in the experiments focusing on 2D quantitative comparisons. The results on the XY, YZ, and XZ planes are presented in Figs. \ref{figS1}, comparing our VTCD approach with other methods, including CycleGAN, DSAR, and Neuroclear. Our method consistently outperformed these baselines across different cell stages (200, 530, and 580 cells), demonstrating enhanced super-resolution (SR) and de-noising capabilities. The results show significant noise reduction and a realistic enhancement of cell membrane outlines, without introducing excessive or artificial details. Notably, our VTCD approach performed well even in the XZ plane, where other methods, such as VTCD+IPG, tended to overemphasize certain features, leading to unrealistic details.

Enlarged viewings are available for detailed comparisons (Fig. 4 and Fig. \ref{figS2}), where improvements in membrane outlines and overall image clarity are apparent. The VTCD method effectively preserves cellular morphology while reducing noise, ensuring that essential structural features are retained. This capability is especially crucial for capturing accurate biological information in fluorescence microscopy data.

\subsection{The 3D Quantitative Demonstration}



We also show a series of 3D quantitative experimental results on dataset NorefZ-2 to further demonstrate the effectiveness of our approach. The comparisons were made between the original noisy images, the outputs from previous methods, and VTCD model (Fig. 3). As seen in the figure, our method was able to reduce noise and restore finer structural details, resulting in processed images' qualities.

\begin{figure}[t]
  \centering
   \includegraphics[width=1.\linewidth]{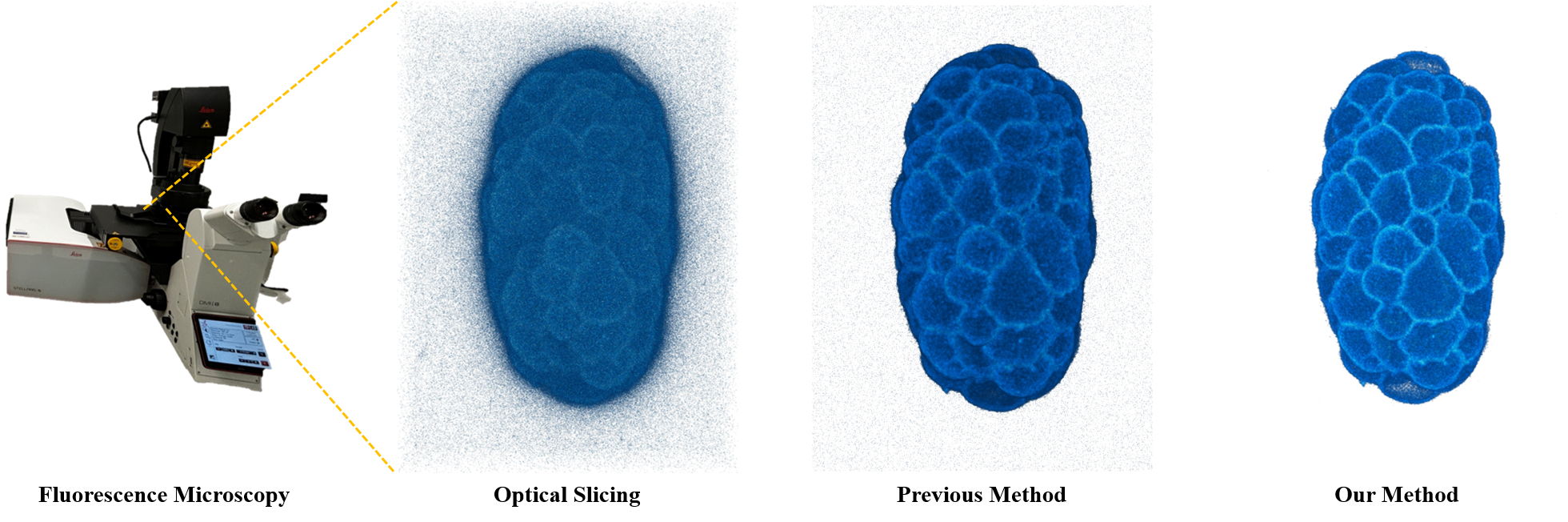}

   \caption{The 3D qualitative comparisons on fluorescence images of original images, previous method result and our method output.}
   \label{figS4}
\end{figure}

In Fig. \ref{figS4}, we provide another set of comparisons highlighting the full workflow from the microscopy equipment (shown on the left) to the processed images. The visual improvements in 3D structural accuracy are evident, with our model achieving clearer outlines and a better representation of the cell boundaries. Unlike other methods that tend to introduce unwanted artifacts during de-noising or resolution enhancement, our VTCD model provides a consistent balance between noise reduction and cell structure preservation.

The rendered 3D outputs demonstrate the efficiency of our cycle-consistent diffusion approach in enhancing fluorescence microscopy data. By reducing the noise while retaining essential morphological features, our method enables more accurate biological interpretation, which is crucial for studying developmental processes at the cellular level. This makes VTCD a robust tool for handling complex 3D fluorescence microscopy data, ensuring high-resolution results even in challenging imaging conditions.

\section{Methodology Supplementary Descriptions}


To help reverse the cell structures in ambiguous FM images and utilize the VTBC's generalization and accuracy, we designed a integrated loss function and train our method in a progressive way.

The training process for our dual cycle-consistent diffusion model, specifically designed for de-noising and super-resolution (SR) in 3D fluorescence microscopy, is structured to address both noise suppression and resolution enhancement. In each iteration, we begin with a sequential training approach where the model is first trained on de-noising tasks, focusing on accurately removing noise while preserving critical features in low-resolution images. In this phase, the model learns noise patterns and low-frequency details, forming a solid foundation for subsequent SR tasks. After stabilizing the de-noising phase, we proceed to train the model for SR, where it learns to recover high-frequency details from downsampled images. Once both tasks are effectively learned independently, we enable dual cycle-consistent training, in which de-noising and SR tasks are jointly optimized. The model employs diffusion-based sampling to iteratively improve image quality by propagating high-quality features across cycles. We also incorporate progressive up-scaling and multi-scale feature discriminators to ensure the SR network captures spatial details while maintaining fidelity across scales, making it particularly effective for 3D fluorescence microscopy data.


To train our VTCD, the basic adversarial losses and the cycle consistent losses are first used to enable the cycle training. With the diffusion forward stage formulated as the targeted slicing of 3D cell volume, the diffusion loss is integrated into the cycle consistency loss by changing the form of the generated process. By the way, to control the generation trajectory, another two losses are formed in the training of the denoising and SR.  The total loss is formulated as follows:

\begin{equation}
    \mathcal{L}_{\text{VTDC}} = \mathcal{L}^{G_A}_{\text{LR $\rightarrow$ HR}} + \mathcal{L}_{\text{HR $\rightarrow$ LR}}^{G_B} + \mathcal{L}_{x_t \rightarrow \hat{x_t}}^{\text{De-noise}} + \mathcal{L}_{\text{XZ/YZ} \rightarrow \text{XY}}^{\text{SR}} 
\end{equation}

Adversarial loss $\mathcal{L}_{\text{LR $\rightarrow$ HR}}^{G_A}$ and $\mathcal{L}_{\text{HR $\rightarrow$ LR}}^{G_B}$ encourages realistic outputs, enabling the model to produce clean, high-resolution images that are indistinguishable from high-quality fluorescence images. Cycle consistent loss for denoising $\mathcal{L}_{x_t \rightarrow \hat{x_t}}^{\text{De-noise}}$ is consist of $\mathcal{L}_{x_t \rightarrow x_0}^{\text{De-noise}} + \mathcal{L}_{TV}(\hat{I})$, while $\mathcal{L}_{TV}(\hat{I})$ is defined as:

\begin{equation}
    \mathcal{L}_{TV}(\hat{I})= \\
    \frac{1}{hwc}\sum_{i,j,k}\sqrt{(\hat{I}_{i,j+1,k}-\hat{I}_{i,j,k})^2+(\hat{I}_{i+1,j,k}-\hat{I}_{i,j,k})^2}
\end{equation}
$\mathcal{L}_{\text{XZ/YZ} \rightarrow \text{XY}}^{\text{SR}}$ is the cycle consistent loss for SR conditioned with $\mathcal{L}_{content}(\hat{I},I;\phi,l)$. In the reverse process of diffusion, to control the generation trajectory, the following losses are used to formulate the conditioning diffusion model. $\mathcal{L}_{content}$ is used in the latent space diffusion model to accurately transform/diffuse the Z-axis images to clear levels and is defined as:

\begin{equation}
    \mathcal{L}_{content}(\hat{I},I;\phi,l)= \\ 
    \frac{1}{h_lw_lc_l}\sqrt{\sum_{i,j,k}(\phi_{i,j,k}^{(l)}(\hat{I}))^2-\phi_{i,j,k}^{(l)}(I)^2)}
\end{equation}
To ensure optimal balance, we dynamically adjust loss weights as the model transitions from denoising-focused to SR-focused stages, refining the fidelity and clarity of the final super-resolved, de-noised outputs.



\section*{Code and Data}

The code is deposited on https://github.com/temporarysharing1688/VTCD and the data is available upon request or after revision. 
